# Electrical Control of 2D Magnetism in Bilayer CrI$_3$

Bevin Huang[1†], Genevieve Clark[2†], Dahlia R. Klein[3†], David MacNeill[3], Efrén Navarro-Moratalla[4], Kyle L. Seyler[1], Nathan Wilson[1], Michael A. McGuire[5], David H. Cobden[1], Di Xiao[6], Wang Yao[7], Pablo Jarillo-Herrero[3*], Xiaodong Xu[1,2*]

[1]Department of Physics, University of Washington, Seattle, Washington 98195, USA
[2]Department of Materials Science and Engineering, University of Washington, Seattle, Washington 98195, USA
[3]Department of Physics, Massachusetts Institute of Technology, Cambridge, Massachusetts 02139, USA
[4]Instituto de Ciencia Molecular, Universidad de Valencia, 46980 Paterna, Spain
[5]Materials Science and Technology Division, Oak Ridge National Laboratory, Oak Ridge, Tennessee, 37831, USA
[6]Department of Physics, Carnegie Mellon University, Pittsburgh, Pennsylvania 15213, USA
[7]Department of Physics and Center of Theoretical and Computational Physics, University of Hong Kong, Hong Kong, China

[†]These authors contributed equally to this work.
[*]Correspondence to: xuxd@uw.edu, pjarillo@mit.edu

**Abstract:** The challenge of controlling magnetism using electric fields raises fundamental questions and addresses technological needs such as low-dissipation magnetic memory[1]. The recently reported two-dimensional (2D) magnets provide a new system for studying this problem owing to their unique magnetic properties[2–8]. For instance, bilayer chromium triiodide (CrI$_3$) behaves as a layered antiferromagnet with a magnetic field-driven metamagnetic transition[2,6]. Here, we demonstrate electrostatic gate control of magnetism in CrI$_3$ bilayers, probed by magneto-optical Kerr effect (MOKE) microscopy. At fixed magnetic fields near the metamagnetic transition, we realize voltage-controlled switching between antiferromagnetic and ferromagnetic states. At zero magnetic field, we demonstrate a time-reversal pair of layered antiferromagnetic states which exhibit spin-layer locking, leading to a remarkable linear dependence of their MOKE signals on gate voltage with opposite slopes. Our results pave the way for exploring new magnetoelectric phenomena and van der Waals spintronics based on 2D materials.

**Main Text:**

Achieving electrical control of magnetism has long been a goal of condensed matter physics and materials science. It not only addresses fundamental aspects of magnetic phenomena and phase transitions[9,10,11], but also has technological importance for developing electrically coupled spintronic devices, such as voltage-controlled magnetic memories with low operation energy[1,12,13]. Previous studies on dilute magnetic semiconductors such as (Ga,Mn)As and (In,Mn)Sb have demonstrated large modulations of the Curie temperatures and coercive fields by altering the magnetic anisotropy and exchange interaction[1,9,14–16]. Recent work on two-dimensional (2D) magnets offers new opportunities to explore the electrical control of magnetic properties[2–4]. In this work, we demonstrate electrical tunability of 2D magnetism using bilayer $CrI_3$ as a model system.

Bilayer $CrI_3$ is a layered antiferromagnet with a Néel temperature of about 45 K[2,6]. As shown in Fig. 1a, spins within each monolayer are aligned ferromagnetically (FM) out-of-plane, while their interlayer coupling is antiferromagnetic (AFM), which results in a vanishing net magnetization. In principle, there are two energetically degenerate AFM ground states which form a time-reversal pair, which we label $\uparrow\downarrow$ and $\downarrow\uparrow$, where the first and second arrows denote the out-of-plane magnetizations in the top and bottom layers, respectively. Previous work has shown that upon increasing the magnetic field in the out-of-plane direction, bilayer $CrI_3$ undergoes a metamagnetic transition from a layered AFM state to an FM state ($\uparrow\uparrow$ or $\downarrow\downarrow$)[2]. Magnetic fields of about 0.6-0.8 T are sufficient to drive the metamagnetic transition, indicating that the interlayer AFM coupling is weak and on the same energy scale. This layered AFM ordering with weak AFM coupling across the van der Waals (vdW) gap presents an outstanding opportunity to realize electrically tunable magnetism.

To explore gate-controlled magnetoelectric effects, we fabricated gated bilayer $CrI_3$ devices. Bulk $CrI_3$ crystals are mechanically exfoliated onto 285-nm $SiO_2$/Si substrates in an argon glove box to obtain bilayer $CrI_3$ flakes (see Methods). The structure of a representative bilayer $CrI_3$ device is shown in Fig. 1b. This device is a vertical stack comprised of a bilayer $CrI_3$ flake and a graphite contact encapsulated between two hexagonal boron nitride (hBN) flakes and a graphite top gate (see Methods)[17]. The hBN flakes protect the bilayer $CrI_3$ from degradation in ambient conditions and also act as dielectric layers for electrostatic doping upon gating. Figure 1c shows a brightfield micrograph of a gated bilayer $CrI_3$ device. The out-of-plane magnetization was then probed either by polar magneto-optical Kerr effect (MOKE) microscopy for devices without a top gate or by reflectance magneto-circular dichroism (RMCD) microscopy for dual-gated devices. Both techniques directly measure the out-of-plane magnetization (see Methods)[18]. A 633 nm HeNe laser focused to a 1 $\mu$m beam spot with 10 $\mu$W of optical power was used for all experiments, which were performed at 15 K unless otherwise specified.

Figure 1d shows a representative RMCD signal of dual-gated bilayer $CrI_3$ (Device 1) as a function of an out-of-plane magnetic field ($\mu_o H$) at zero gate voltage. The orange curve corresponds to sweeping $\mu_o H$ from a negative to a positive value and the green curve shows the time-reversed process. For magnetic fields smaller than 0.8 T, the RMCD signal nearly vanishes, corresponding to the layered AFM states, either $\uparrow\downarrow$ or $\downarrow\uparrow$. However, unlike the as-exfoliated bilayer samples measured without vdW assembly[2], there is a remnant RMCD signal, implying that the net magnetization is not completely suppressed in the AFM configurations. We often observe this remnant RMCD signal in bilayer devices that have been through the vdW assembly process. This could be caused by breaking out-of-plane symmetry of the bilayer during device fabrication such that the magnetizations from the two layers do not exactly cancel. At a critical field, $\mu_oH_c = \pm 0.8$ T, we observe a sharp jump in the RMCD signal from near-zero up to 3% and this increased signal is spatially homogeneous across the entire device (Fig. S1). This is consistent with the previous report for as-exfoliated bilayer CrI$_3$, implying that an external magnetic field drives a metamagnetic transition in which the bilayer switches from the layered AFM states (↑↓ or ↓↑) to an FM state (↑↑ or ↓↓)[2].

Remarkably, the metamagnetic transition is highly controllable by electrostatic gating. We performed MOKE measurements on a second device fabricated with a single gate (Device 2) while varying both the magnetic field and gate voltage. In this device, we apply the gate voltage through a SiO$_2$/Si back gate while leaving the bilayer CrI$_3$ uncovered to minimize the impact of device fabrication on the CrI$_3$ bilayer. Figure 2a shows a plot of MOKE intensity as a function of the back gate voltage ($V_{bg}$) and $\mu_oH$ (sweeping down). In this magnetic phase diagram, the light color represents low MOKE signal arising from a layered AFM state, while the darker red and blue colors represent the two FM states with strong MOKE signals of opposite signs. Based on this phase diagram, the phase boundary, *i.e.* the critical field for the metamagnetic transition, strongly depends on $V_{bg}$.

In Fig. 2b, we look closer at the metamagnetic transition by comparing the magnetic field dependence at $V_{bg} = 0$ V and $\pm 50$ V (horizontal line cuts of the phase map of Fig. 2a). The critical field $\mu_oH_c$ has shifted by roughly 0.2 T, from 0.7 T at -50 V to 0.5 T at 50 V. Considering that the critical field is about 0.6 T for $V_{bg} = 0$ V, this corresponds to tuning the critical field by around 30% via electrostatic gating. The shift in $\mu_oH_c$ is bidirectional and linear; that is, when a positive gate voltage is applied, the absolute value of the critical field is decreased, and vice versa for negative gate voltages.

Due to this strong dependence of the critical field on the gate voltage, we can then realize an electrically controlled transition between the layered AFM and FM phases when the magnetic field is fixed near $\mu_oH_c$. Figure 2c illustrates such an electrically controlled magnetic phase transition at three selected magnetic fields, corresponding to vertical line cuts on the phase map in Fig. 2a. The MOKE signal starts near zero at negative gate voltages, corresponding to AFM ordering. As the voltage increases to a critical value, the MOKE signal quickly rises and reaches a saturated value, demonstrating electrical switching to the FM state. For the three selected magnetic fields ranging from 0.58 T to 0.62 T, the required voltage to initiate the transition from the layered AFM states to the FM ↑↑ state shifts towards positive gate voltage with decreasing applied magnetic field.

In single-gated devices, an applied gate voltage introduces both an interlayer bias and layer-dependent electrostatic doping. To determine which of the two effects dominates the gate-tunable magnetic phase transition, we investigated a dual-gated device to enable independent control over the doping and electric field. Under application of a single gate, the dual-gated device behaves similarly to Device 2 (Figs. S2 & S3). After identifying the critical field for the metamagnetic transition, we fixed the magnetic field near the transition and measured the RMCD signal as a function of both gates.

Figure 3a shows the RMCD intensity plot as a function of both top and back gate voltages at $\mu_oH = 0.78$ T. The graphite top gate was swept between $\pm 6$ V and the silicon back gate between $\pm 15$ V. The red region on the right is the large, positive signal from the ↑↑ state while the pink region on the left is the much smaller signal from the layered AFM state. Using a parallel-plate capacitor model[19–21] we found the constant-doping and constant-field contours on the phase map (Methods). The metamagnetic transition, *i.e.* the boundary between the red and pink regions, lies nearly parallel to the constant-doping contours as indicated by the dashed line in Fig. 3a (for the time-reversal counterpart at negative magnetic fields, see Fig. S4). To further confirm this, we performed magnetic field sweeps of the RMCD signal at several doping levels with the same displacement field (Fig. 3b), and at several displacement fields with the same doping level (Fig. 3c). These clearly show that the doping level dominantly affects the critical field. This is not surprising since in other magnetic systems, doping leads to modulation of orbital occupation and magnetic anisotropy[1,9,14,15,22–24], and thus a change in the magnetism. Our observation may stimulate theoretical investigations of whether doping in 2D magnets introduces similar modulation effects.

In addition to gate-voltage tuning of the metamagnetic transition, we further realize electrical control of the layered AFM states at zero magnetic field. We first prepare the system by starting in the ↑↑ state with an applied magnetic field of 1 T, and then sweeping the field back to zero (Initialization 1, inset Fig. 4a). Remaining at zero external magnetic field, we sweep the back gate voltage and observe a MOKE signal linearly dependent on the voltage[25] with a slope of -0.04 mrad/V, shown in Fig. 4a. Similarly, we prepare the bilayer to its time-reversal state by starting in the ↓↓ state with a magnetic field of -1 T and sweeping the field back to zero (Initialization 2, inset Fig. 4b). In Fig. 4b, the MOKE signal also exhibits a linear dependence on the gate voltage, but with an opposite slope of 0.04 mrad/V. These results demonstrate that the two initialization processes lead to two distinct layered AFM states, ↑↓ and ↓↑, which are time-reversed pairs, and that we have full electrical tunability over the AFM states even at zero magnetic field.

This observed magnetoelectric coupling in the layered AFM states can be explained by the spin-layer locking effect. For an AFM state at zero field, the spin (magnetization) orientation is locked to the layer pseudospin that labels the geometric top and bottom layers (Fig. 1a). A direct consequence of this spin-layer locking effect is that the creation of a layer polarization is accompanied by a net magnetization[26,27]. In the AFM configuration, the opposite spin orientations of the two layers quench the interlayer hopping and hence the layer hybridization[20,26]. The application of an electrostatic gate voltage then creates a layer polarization of the carriers, resulting in net magnetizations of opposite signs in the ↑↓ and ↓↑ states. The layer polarization and hence net magnetization has a linear dependence on the gate voltage, consistent with the observed gate dependence of the MOKE signal in Fig. 4.

Furthermore, this allows us to identify and controllably access the two energetically degenerate AFM states. The application of a positive gate voltage results in a larger electron density in the bottom layer than in the top layer. If the bilayer is in the ↑↓ state, the net magnetization then points down with a negative slope of the MOKE signal as a function of gate voltage. The ↓↑ state will then have a positive slope in the gate-dependent MOKE signal. Although the device assembly process may already break the degeneracy between the ↑↓ and ↓↑ states, the signs of the gate-dependent MOKE slopes are nevertheless fingerprints for distinguishing between the two AFM states. Based on Figs. 4a and 4b, we can definitively determine that by using Initialization 1, the ↑↑ state transitions to the ↑↓ state, and by using Initialization 2, the ↓↓ state transitions to the ↓↑ state.

In summary, we have demonstrated two forms of electrical control of magnetism in a van der Waals layered antiferromagnetic insulator, bilayer CrI$_3$. First, in an applied magnetic field, the critical field for the metamagnetic transition can be tuned by up to 30% by applying a gate voltage, allowing a gate voltage driven phase transition from a layered AFM phase to an FM phase. Dual gate studies show that the mechanism is dominated by electrostatic doping. Second, at zero magnetic field, the net magnetization can be continuously tuned by gate voltage. The two antiferromagnetic states, ↑↓ and ↓↑, show the same linear dependence of the MOKE signal on gate voltage but with opposite signs, which provides an experimental measure to distinguish these two degenerate magnetic states. Our work shows that van der Waals magnets provide a new system for exploring magnetoelectric effects and their potential applications in gate-tunable spintronics.

**References:**

1. Matsukura, F., Tokura, Y. & Ohno, H. Control of magnetism by electric fields. *Nat. Nanotechnol.* **10,** 209–220 (2015).
2. Huang, B. *et al.* Layer-dependent ferromagnetism in a van der Waals crystal down to the monolayer limit. *Nature* **546,** 270–273 (2017).
3. Gong, C. *et al.* Discovery of intrinsic ferromagnetism in two-dimensional van der Waals crystals. *Nature* **546,** 265–269 (2017).
4. Xing, W. *et al.* Electric field effect in multilayer $Cr_2Ge_2Te_6$ : a ferromagnetic two-dimensional material. *2D Mater.* **4,** 24009 (2017).
5. Lee, J. U. *et al.* Ising-type magnetic ordering in atomically thin $FePS_3$. *Nano Lett.* **16,** 7433–7438 (2016).
6. Seyler, K. L. *et al.* Ligand-field helical luminescence in a 2D ferromagnetic insulator. *Nat. Phys.* http://dx.doi.org/10.1038/s41567-017-0006-7 (2017).
7. Wang, X. *et al.* Raman spectroscopy of atomically thin two-dimensional magnetic iron phosphorus trisulfide ($FePS_3$) crystals. *2D Mater.* **3,** 31009 (2016).
8. McGuire, M. A., Dixit, H., Cooper, V. R. & Sales, B. C. Coupling of crystal structure and magnetism in the layered, ferromagnetic insulator $CrI_3$. *Chem. Mater.* **27,** 612–620 (2015).
9. Chiba, D. *et al.* Magnetization vector manipulation by electric fields. *Nature* **455,** 515–518 (2008).
10. Chiba, D. *et al.* Electrical control of the ferromagnetic phase transition in cobalt at room temperature. *Nat. Mater.* **10,** 853–856 (2011).
11. Eerenstein, W., Mathur, N. D. & Scott, J. F. Multiferroic and magnetoelectric materials. *Nature* **442,** 759–765 (2006).
12. Prinz, G. A. Magnetoelectronics. *Science* **282,** 1660–1663 (1998).
13. Chu, Y. H. *et al.* Electric-field control of local ferromagnetism using a magnetoelectric multiferroic. *Nat. Mater.* **7,** 478–482 (2008).
14. Ohno, H. *et al.* Electric-field control of ferromagnetism. *Nature* **408,** 944–947 (2000).
15. Dietl, T., Ohno, H., Matsukura, F., Cibert, J. & Ferrand, D. Zener model description of ferromagnetism in zinc-blende magnetic semiconductors. *Science* **287,** 1019–1022 (2000).
16. MacDonald, A. H., Schiffer, P. & Samarth, N. Ferromagnetic semiconductors: moving beyond (Ga,Mn)As. *Nat. Mater.* **4,** 195–202 (2005).
17. Wang, L. *et al.* One-dimensional electrical contact to a two-dimensional material. *Science* **342,** 614–617 (2013).
18. Sato, K. Measurement of magneto-optical Kerr effect using piezo-birefringent modulator. *Jpn. J. Appl. Phys.* **20,** 2403–2409 (1981).

19. Zhang, Y. *et al.* Direct observation of a widely tunable bandgap in bilayer graphene. *Nature* **459,** 820–823 (2009).
20. Taychatanapat, T., Watanabe, K., Taniguchi, T. & Jarillo-Herrero, P. Quantum Hall effect and Landau-level crossing of Dirac fermions in trilayer graphene. *Nat. Phys.* **7,** 621–625 (2011).
21. Dean, C. R. *et al.* Boron nitride substrates for high-quality graphene electronics. *Nat. Nanotechnol.* **5,** 722–726 (2010).
22. Weisheit, M. *et al.* Electric field-induced modification of magnetism in thin-film ferromagnets. *Science* **315,** 349–351 (2007).
23. Duan, C. G. *et al.* Surface magnetoelectric effect in ferromagnetic metal films. *Phys. Rev. Lett.* **101,** 137201 (2008).
24. Nakamura, K. *et al.* Giant modification of the magnetocrystalline anisotropy in transition-metal monolayers by an external electric field. *Phys. Rev. Lett.* **102,** 187201 (2009).
25. Sivadas, N., Okamoto, S. & Xiao, D. Gate-controllable magneto-optic Kerr effect in layered collinear antiferromagnets. *Phys. Rev. Lett.* **117,** 267203 (2016).
26. Gong, Z. *et al.* Magnetoelectric effects and valley-controlled spin quantum gates in transition metal dichalcogenide bilayers. *Nat. Commun.* **4,** 2053 (2013).
27. Jones, A. M. *et al.* Spin-layer locking effects in optical orientation of exciton spin in bilayer $WSe_2$. *Nat. Phys.* **10,** 130–134 (2014).

**Acknowledgements:** Work at the University of Washington was mainly supported by the Department of Energy, Basic Energy Sciences, Materials Sciences and Engineering Division (DE-SC0012509), and University of Washington Innovation Award. Work at MIT has been supported by the Center for Integrated Quantum Materials under NSF Grant DMR-1231319 as well as the Gordon and Betty Moore Foundation's EPiQS Initiative through Grant GBMF4541 to PJH. DRK was funded in part by a QuantEmX grant from ICAM and the Gordon and Betty Moore Foundation through Grant GBMF5305 and from the NSF Graduate Research Fellowship Program (GRFP) under Grant 1122374. Device fabrication has been partly supported by the Center for Excitonics, an Energy Frontier Research Center funded by the US Department of Energy (DOE), Office of Science, Office of Basic Energy Sciences under Award Number DESC0001088. DC's contribution is supported by DE-SC0002197. Work at CMU is also supported by DOE BES DE-SC0012509. WY is supported by the Croucher Foundation (Croucher Innovation Award), and the HKU ORA. Work at ORNL (MAM) was supported by the US Department of Energy, Office of Science, Basic Energy Sciences, Materials Sciences and Engineering Division. DX acknowledges the support of a Cottrell Scholar Award. XX acknowledges the support from the State of Washington funded Clean Energy Institute and from the Boeing Distinguished Professorship in Physics.

**Author Contributions:** XX and PJH supervised the project. ENM and MAM synthesized and characterized the bulk $CrI_3$ crystals. DRK fabricated the devices, assisted by DM, GC, and BH. GC built the setup with the help from BH and NW. BH and GC performed the MOKE measurements, assisted by KLS and DRK. WY and DX provided the theoretical support. All authors contributed to the paper writing and discussed the results.

## Methods:

### Device fabrication

Bulk $CrI_3$ crystals were grown by the direct vapor transport method as previously reported[2,8]. All fabrication using these crystals was performed in a glove box with argon atmosphere. The crystals were mechanically exfoliated onto either 285 nm $SiO_2$/Si substrates for dual-gated devices (*e.g.* Device 1) or the commercially available viscoelastic polymer Gel Film from Gel-Pak (*e.g.* Device 2). Bilayer monocrystalline flakes were identified by their optical contrast relative to the substrate.

The van der Waals assembly of the stack for a dual-gated device was performed in the glove box using the dry transfer polymer technique with a poly(bisphenol A carbonate) film placed over a polydimethylsiloxane square serving as our stamp[17]. The flakes were picked up sequentially: graphite top gate, top hexagonal boron nitride, graphite contact, bilayer $CrI_3$, bottom hexagonal boron nitride.

Metallic Cr/Pd contacts on 285 nm $SiO_2$/Si substrates were prepared using standard electron beam lithography and bilayer poly(methyl methacrylate) resist fabrication techniques. Each chip was fixed to a DIP socket with silver paint and wire bonded to the socket.

The completed dual-gated stacks were transferred on top of the Cr/Pd contacts and released by heating to 170ºC. The non-encapsulated devices were fabricated by transferring the as-exfoliated bilayer $CrI_3$ flake on Gel Film directly onto the metallic contacts. Finally, each socket was sealed with a glass coverslip using Crystalbond under inert conditions to permit optical measurements.

### MOKE and RMCD microscopy

Materials that exhibit an out-of-plane magnetization, **M**, may also display magnetic circular birefringence (MCB) and magnetic circular dichroism (MCD). Both MCB and MCD accrue a difference in the phase and amplitude, respectively, between right- and left-circularly polarized (RCP, LCP) light that vary as a function of **M**. When linearly polarized light, an equal superposition of RCP and LCP, is normally incident and reflects off the magnetized material, the phase difference between RCP and LCP leads to a rotation of the linear polarization through an angle, $\theta_K$, from the magneto-optical Kerr effect (MOKE) and induces ellipticity through reflective magnetic circular dichroism (RMCD).

Both MOKE and RMCD measurements were performed in a closed-cycle helium cryostat with a base temperature of 15 K. A superconducting solenoidal magnet allowed fields of up to 7 T in the Faraday geometry to be applied on all devices. The AC lock-in measurement technique used to measure the MOKE and RMCD signal follows closely to previous characterizations of the magnetic order in atomically thin $CrI_3$. For RMCD measurements, the analyzing polarizer was removed, and the lock-in frequency was set to the fundamental frequency of the photoelastic modulator (PEM) rather than its second harmonic[2,18].

**Calculation of constant doping line**

Following the methodology from Zhang *et al.* and Taychatanapat *et al.*, our dual-gated bilayer CrI$_3$ device can be approximated as a parallel-plate capacitor to obtain the constant doping ($n$) and constant electric displacement field (**D**) contours for the dual gate phase map[19,20]. The amount of doping, $n$, in the device is then:

$$n = C_{BG}(V_{BG} - V_{BG}^D) + C_{TG}(V_{TG} - V_{TG}^D) \qquad (1)$$

where $C$ is the areal capacitance, $V_{BG}$ and $V_{TG}$ are the back gate and top gate voltages, respectively, and $V_{BG}^D$ and $V_{TG}^D$ the charge neutral point voltages. Since the aim of this calculation is to show whether the metamagnetic transition shown in Fig. 3a arises from a change in $n$ or a change in **D**, we can find one such constant-doping contour by setting $n = V_{TG} = V_{BG} = 0$, as these terms only offset the contour by an arbitrary constant. The areal capacitance is given as: $C = \varepsilon/d$, where $\varepsilon$ is the permittivity of the dielectric layer and $d$ is the thickness of the dielectric layer. Including the bottom hBN encapsulation layer with the SiO$_2$ dielectric layer in the bottom areal capacitance, equation (1) for Device 1 simplifies to:

$$V_{TG} = -\frac{d_{hBN}}{\frac{\varepsilon_{hBN}}{\varepsilon_{SiO_2}}d_{SiO_2} + d_{hBN}} V_{BG} \qquad (2)$$

The out-of-plane static dielectric constants of hBN and SiO$_2$, $\varepsilon_{hBN}$ and $\varepsilon_{SiO2}$ respectively, are 4 and 3.9[21]. The device sits on top of 285 nm SiO$_2$ substrate, and both top and bottom hBN flakes were 30 nm. Similarly, we can also obtain the zero-displacement field contour by starting with the general expression for the applied displacement field:

$$\mathbf{D} = C_{BG}(V_{BG} - V_{BG}^D) - C_{TG}(V_{TG} - V_{TG}^D) \qquad (3)$$

and removing the same terms as in obtaining Equation (2):

$$V_{TG} = +\frac{d_{hBN}}{\frac{\varepsilon_{hBN}}{\varepsilon_{SiO_2}}d_{SiO_2} + d_{hBN}} V_{BG} \qquad (4)$$

Figures:

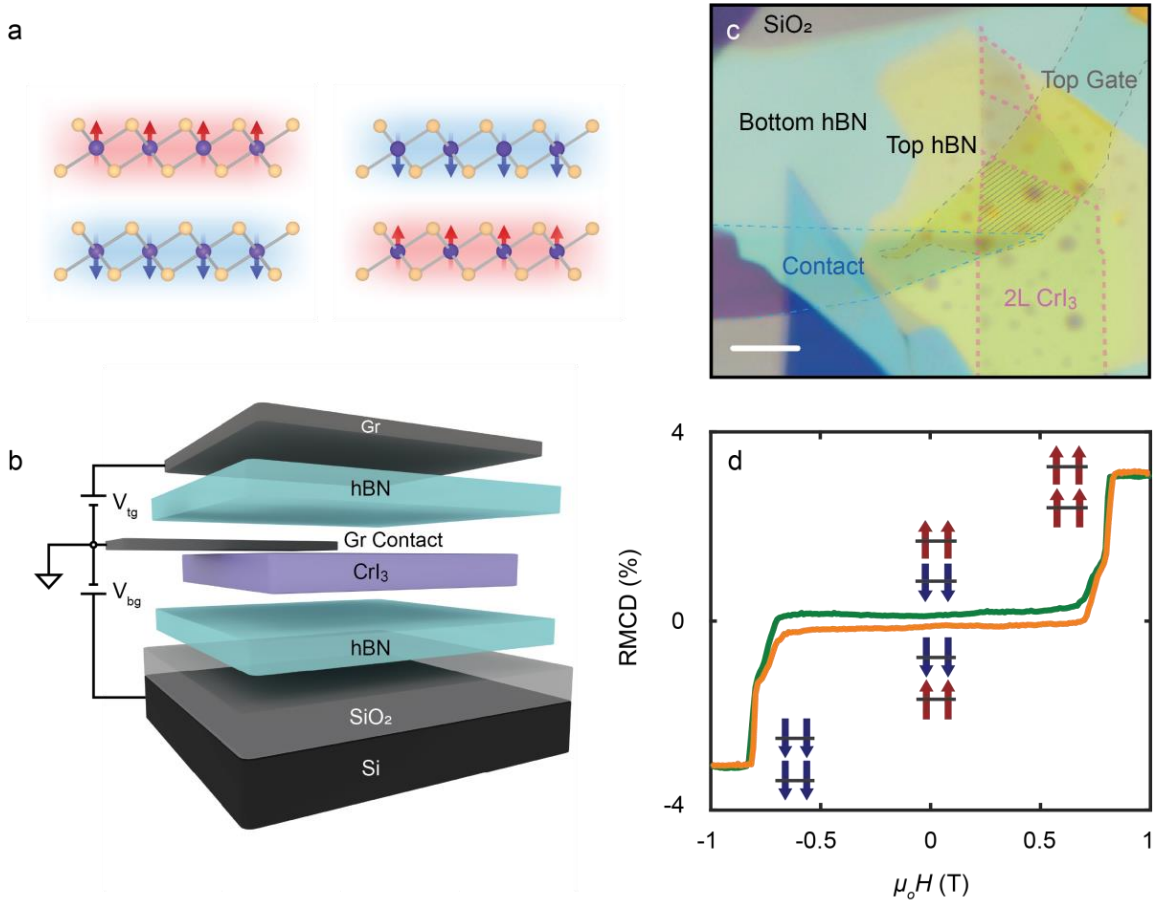

**Figure 1 | Bilayer CrI$_3$ spin ordering, gated device structure, and RMCD without gating.** (a) Magnetic ordering in bilayer CrI$_3$ in the two energetically degenerate antiferromagnetic ground states, labeled ↑↓ (left) and ↓↑ (right) in the text, when cooled below the Néel temperature. Cr$^{3+}$ and I$^-$ ions are represented with purple and orange balls, respectively. (b) Schematic of a dual-gated bilayer CrI$_3$ device fabricated by van der Waals assembly. (c) False-color optical micrograph of a representative device (Device 1). The scale bar is 5 $\mu$m. A small four-layer CrI$_3$ piece (thin pink dashed line) interconnects the two bilayer (2L) sections of the flake (thick pink dashed line). (d) RMCD signal of a bilayer CrI$_3$ device (Device 1) as a function of perpendicular magnetic field at zero gate voltage.

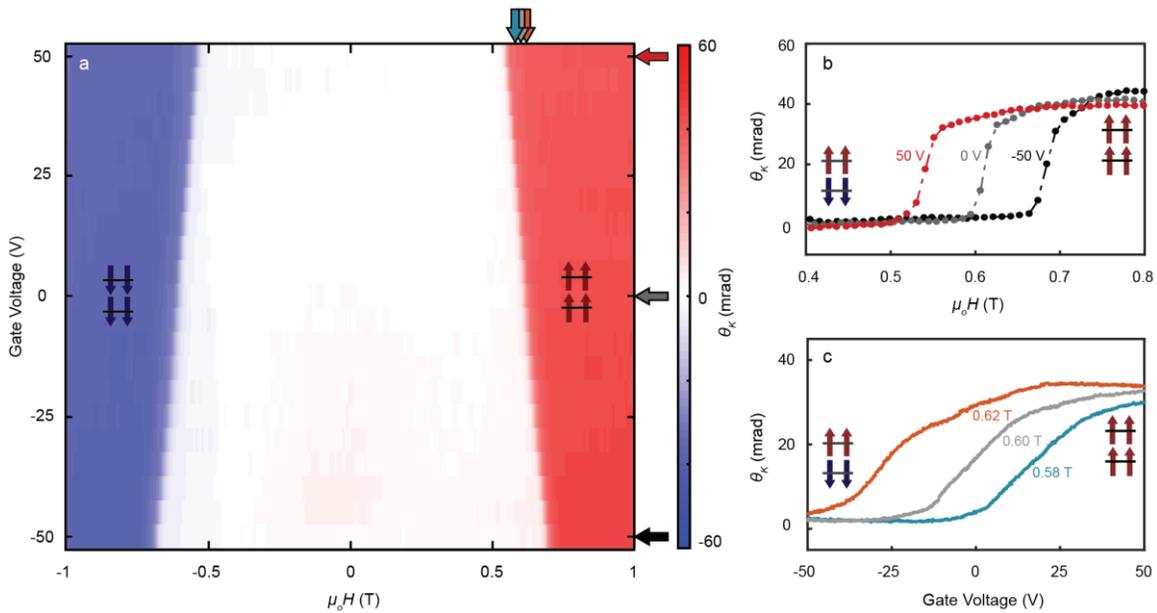

**Figure 2 | MOKE signal vs. gate voltage and applied magnetic field. (a)** Intensity of the polar MOKE signal of a non-encapsulated bilayer CrI$_3$ device (Device 2) as a function of both gate voltage and applied magnetic field. **(b)** Selected horizontal line cuts of (a) demonstrating the gated induced change in critical field of the metamagnetic transition at $V_{bg}$ = -50 V (black), 0 V (gray), and 50 V (red). **(c)** Gate induced transition from layered antiferromagnetic to ferromagnetic states at selected $\mu_0 H$ = 0.58 T (blue), 0.60 T (gray), and 0.62 T (orange).

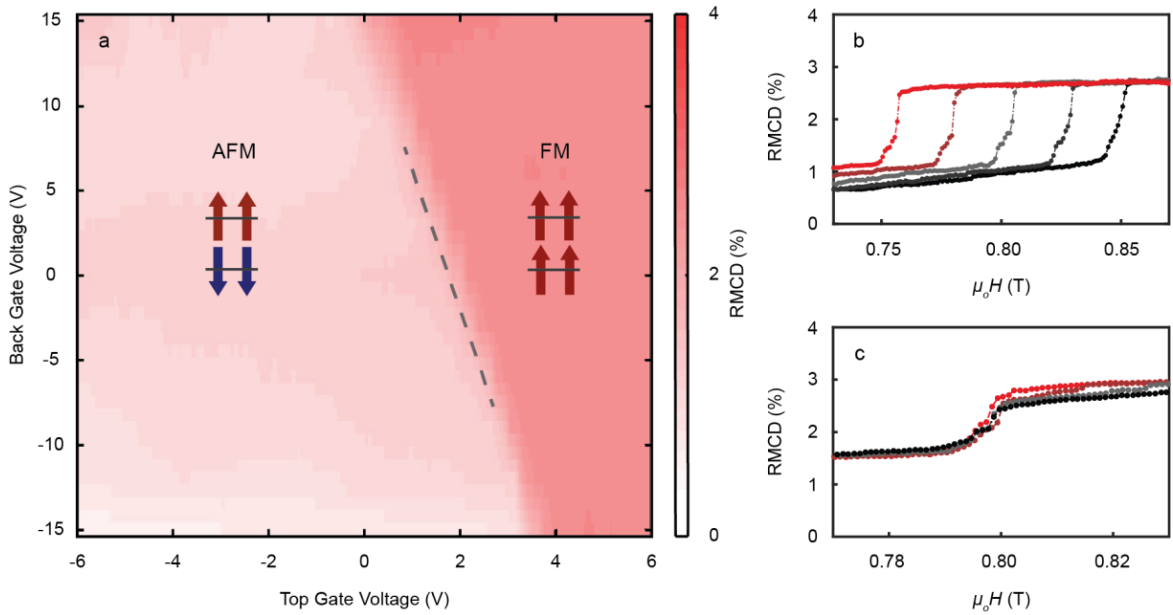

**Figure 3 | Origin of voltage-controlled metamagnetism.** (**a**) RMCD signal of a dual-gated device (Device 1) when sweeping both the graphite top gate and silicon back gate. The pink region reflects negligible RMCD signal, corresponding to the layered antiferromagnetic (AFM) states. The red region corresponds to the ferromagnetic ↑↑ state. The boundary between them is parallel to a constant doping contour (gray dashed line). (**b**) RMCD signal of the same device as a function of applied magnetic field at fixed doping levels from 0 cm$^{-2}$ (black) to 4.4×10$^{12}$ cm$^{-2}$ (red) for zero displacement field. (**c**) RMCD signal as a function of applied magnetic field at several displacement fields ranging from 0 V/nm (red) to 0.6 V/nm (black) for zero doping. The metamagnetic transition is clearly insensitive to changes in the displacement field but strongly dependent on the doping level.

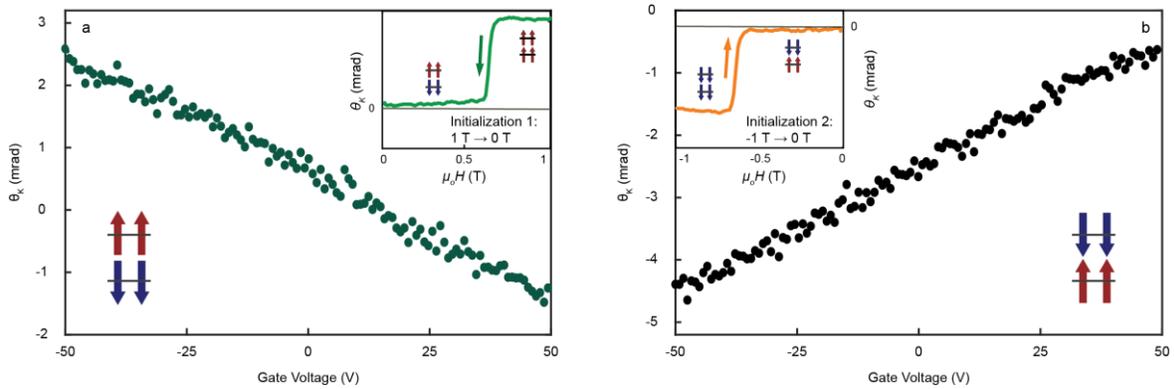

**Figure 4 | Gate-voltage induced MOKE of layered antiferromagnetic states at zero magnetic field. (a)** Gate-dependent MOKE signal of a non-encapsulated bilayer $CrI_3$ device (Device 2) prepared in the ↑↓ state following the initialization depicted in the inset (see text for details). **(b)** Equivalent measurement performed on the device prepared in the ↓↑ state by the initialization process in the inset. The opposite slopes of the MOKE signal between ↑↓ and ↓↑ highlight the spin-layer locking effect with gated-induced MOKE. Insets show the initialization processes for preparing antiferromagnetic states at zero field (discussed in the text).

# Supplementary Information for

# Electrical Control of 2D Magnetism in Bilayer CrI$_3$

Bevin Huang[1†], Genevieve Clark[2†], Dahlia R. Klein[3†], David MacNeill[3], Efrén Navarro-Moratalla[4], Kyle Seyler[1], Nathan Wilson[1], Michael A. McGuire[5], David H. Cobden[1], Di Xiao[6], Wang Yao[7], Pablo Jarillo-Herrero[3*], Xiaodong Xu[1,2*]

[1]Department of Physics, University of Washington, Seattle, Washington 98195, USA
[2]Department of Materials Science and Engineering, University of Washington, Seattle, Washington 98195, USA
[3]Department of Physics, Massachusetts Institute of Technology, Cambridge, Massachusetts 02139, USA
[4]Instituto de Ciencia Molecular, Universidad de Valencia, 46980 Paterna, Spain
[5]Materials Science and Technology Division, Oak Ridge National Laboratory, Oak Ridge, Tennessee, 37831, USA
[6]Department of Physics, Carnegie Mellon University, Pittsburgh, Pennsylvania 15213, USA
[7]Department of Physics and Center of Theoretical and Computational Physics, University of Hong Kong, Hong Kong, China

[†]These authors contributed equally to this work.
[*]Correspondence to: xuxd@uw.edu, pjarillo@mit.edu

**Contents:**

**Fig. S1:** RMCD maps at fixed magnetic fields near the metamagnetic transition

**Fig. S2:** RMCD maps at fixed top gate voltages

**Fig. S3:** Single-gate RMCD measurements vs. gate-voltage and magnetic field on Device 1

**Fig. S4:** Dual-gate phase map at the negative metamagnetic transition

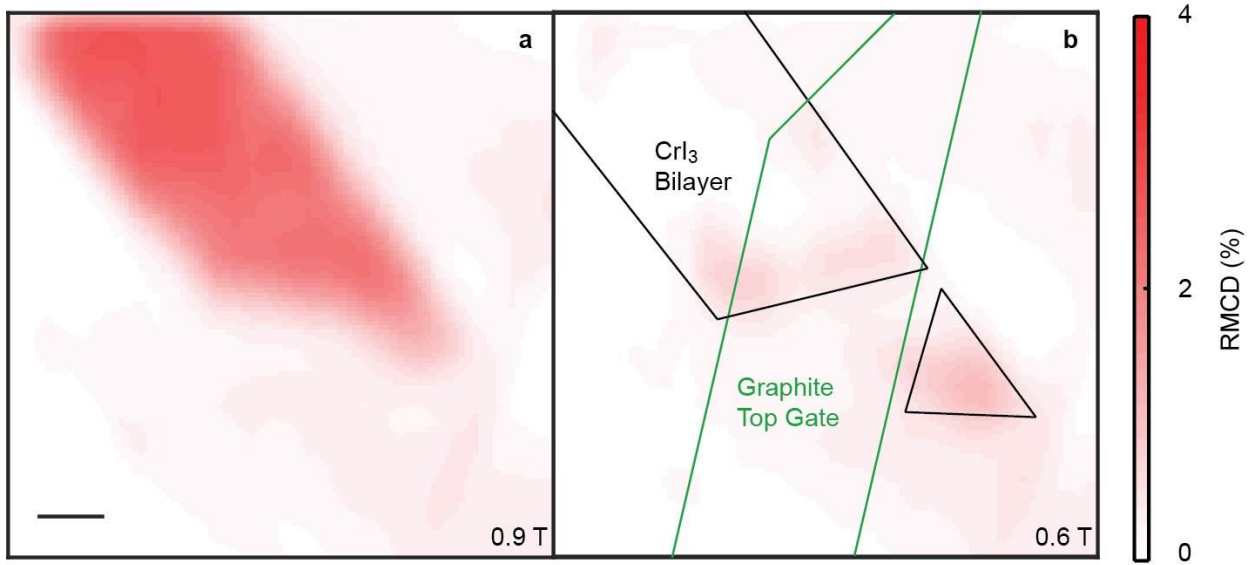

**Figure S1 | RMCD maps at fixed magnetic fields near the metamagnetic transition.** RMCD map of Device 1 in the absence of gate voltage with an applied magnetic field values of **(a)** 0.9 T and **(b)** 0.6 T. These maps show the metamagnetic transition from the fully spin-polarized ferromagnetic state ↑↑ (red) to the layered antiferromagnetic state ↑↓ (white) near the critical field regime. The ferromagnetism beyond the metamagnetic transition is homogenous across the entire bilayer. The scale bar is 5 $\mu$m.

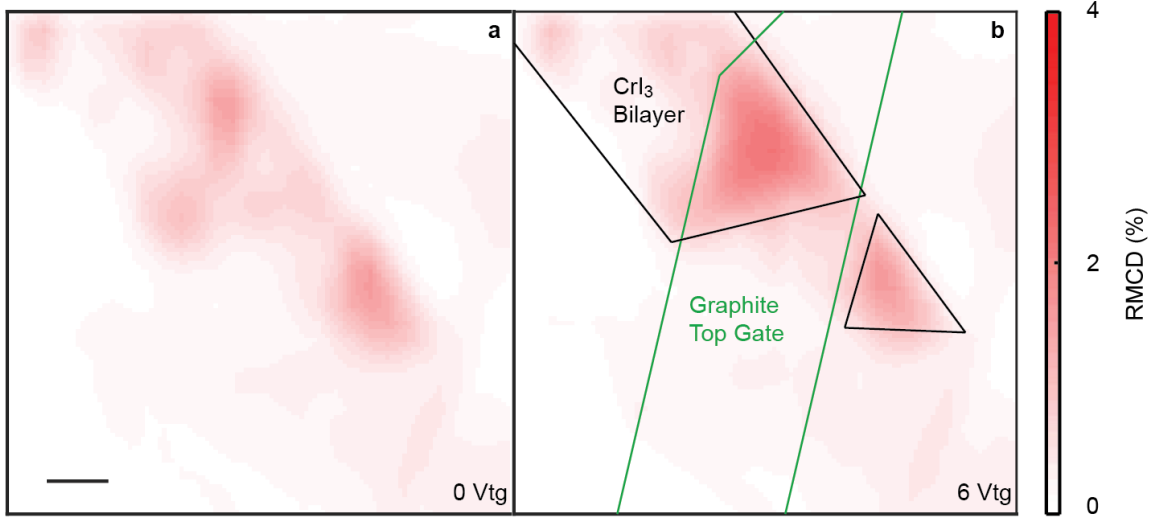

**Figure S2 | RMCD maps at fixed top gate voltages. (a)** RMCD map of Device 1 with an applied magnetic field of 0.8 T in the absence of gating. The scale bar is 5 $\mu$m. **(b)** RMCD map of the same region in (a) at a magnetic field of 0.8 T with zero back gate voltage and 6 V applied to the graphite top gate. Similar to the gate voltage sweeps of the RMCD signal in Fig. 2c, we are able to electrically switch the antiferromagnetic state (light pink) to the ferromagnetic ↑↑ state (red) by applying a positive top gate voltage on the bilayer CrI$_3$ flake. This also illustrates the homogeneity of the gating across the portion of the bilayer flake under the graphite top gate.

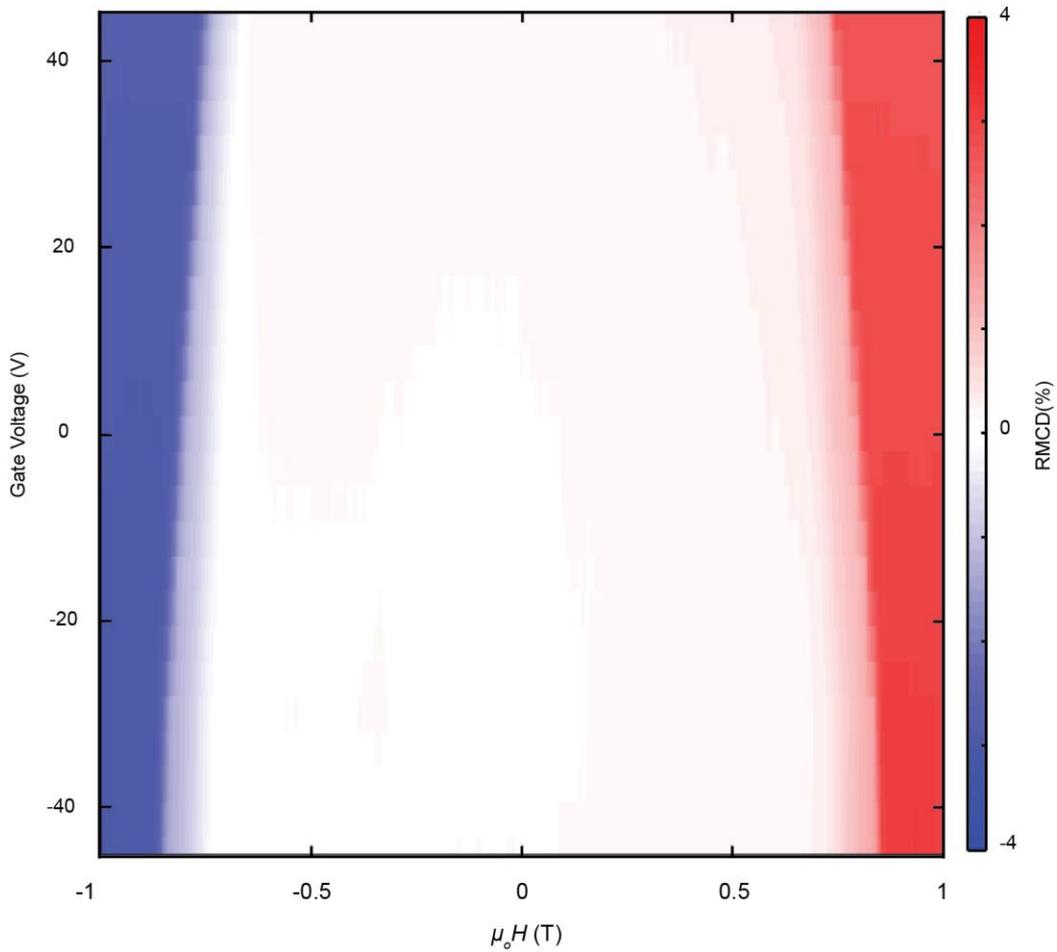

**Figure S3 | Single-gate RMCD measurements vs. gate-voltage and magnetic field on Device 1.** RMCD intensity as a function of applied back gate voltage and magnetic field. Maintaining the same convention as in Fig. 2a, the bright red region corresponds to a large positive RMCD signal associated with the ferromagnetic ↑↑ state, while the dark blue region shows large negative RMCD signal that corresponds to the ↓↓ state. The light regions in between the blue and red regions are instances where the RMCD signal is near-zero and when the bilayer is in a layered-antiferromagnetic state. Consistent with Device 2, the critical field of the metamagnetic transition lowers as a positive gate voltage is applied and vice versa for negative gate voltages.

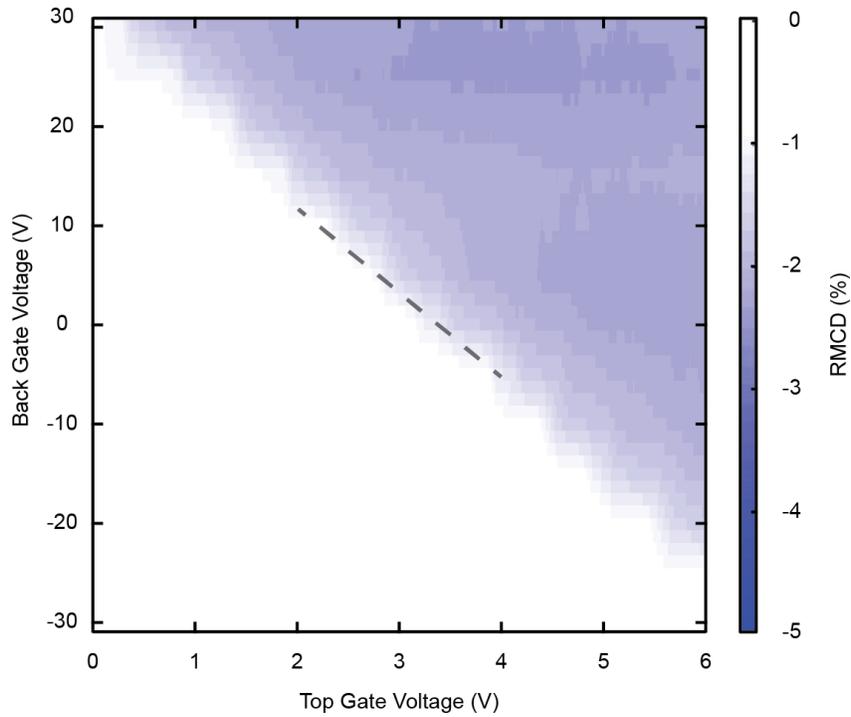

**Figure S4 | Dual-gate phase map at the negative metamagnetic transition.** Phase map of the RMCD signal as a function of both the top gate and back gate voltages in Device 1. The blue region represents a strong, negative RMCD signal, *i.e.* when the bilayer is in the ferromagngetic ↓↓ state, whereas the white region represents the absence of RMCD signal of the antiferromagnetic ↓↑ state. The gray dashed line shows a constant doping contour. Similar to the phase boundary seen in the positive metamagnetic transition dual-gate phase map (Fig. 3a), the negative metamagnetic transition also reveals doping to be the dominant factor in tuning the magnetic states at large negative fields.